\documentclass[sigconf]{acmart}
\usepackage{enumitem}

\AtBeginDocument{%
  }

\renewcommand\footnotetextcopyrightpermission[1]{}

\begin{document}

\title[Toward Bounding Coding-Agent Autonomy]
{Toward Bounding Coding-Agent Autonomy in Software Maintenance: A Pilot Study of Boundary Contracts}

\author{Heeyong Eun}
\affiliation{%
  \institution{KPMG Korea}
  \city{Seoul}
  \country{South Korea}
}
\email{heeyonge@gmail.com}

\author{Seungyeon Chung}
\affiliation{%
  \institution{KPMG Korea}
  \city{Seoul}
  \country{South Korea}
}
\email{seungyeonchung@kr.kpmg.com}

\author{Hyejee Park}
\affiliation{%
  \institution{KPMG Korea}
  \city{Seoul}
  \country{South Korea}
}
\email{hyejeepark@kr.kpmg.com}

\author{Youngeun Lee}
\affiliation{%
  \institution{KPMG Korea}
  \city{Seoul}
  \country{South Korea}
}
\email{ylee143@kr.kpmg.com}

\author{Sukhyun Hong}
\affiliation{%
  \institution{KPMG Korea}
  \city{Seoul}
  \country{South Korea}
}
\email{shong31@kr.kpmg.com}

\author{Hyewon Jung}
\affiliation{%
  \institution{KPMG Korea}
  \city{Seoul}
  \country{South Korea}
}
\email{hjung22@kr.kpmg.com}

\renewcommand{\shortauthors}{Eun et al.}

\begin{abstract}
As coding agents gain greater autonomy to modify software repositories, relevant context alone does not define their operational authority. An artifact may be relevant to a maintenance task while being inappropriate to edit; conversely, another artifact may be required even when it is not explicitly mentioned in the request. We explore \emph{Boundary Contracts}, a machine-readable abstraction that makes task-specific artifact roles---editable, inspect-only, required, and protected---explicit before repository modification begins.

We conduct a pilot study on 40 maintenance tasks from a real enterprise repository, comparing the same Codex-based coding agent with and without Boundary Contract information. The boundary-informed condition improves macro edit-selection F1 from 0.440 to 0.682, required-behavior coverage from 0.429 to 0.673, and the number of strictly acceptable patches from 9 of 40 to 18 of 40. To probe whether these gains are explained only by improved file localization, we additionally analyze 21 tasks in which both conditions modify at least one identical file. Differences remain in edit selectivity and patch outcomes, providing initial evidence that explicit artifact roles may influence maintenance behavior beyond localization alone.

This pilot does not establish the general effectiveness of Boundary Contracts. Rather, it provides initial evidence for a broader research question: how should the operational boundaries of coding agents be generated, validated, revised, and ultimately enforced as repository-level maintenance becomes increasingly autonomous?
\end{abstract}


\maketitle

%section1

\section{Introduction}
\label{sec:introduction}

Large language model (LLM)-based coding agents are increasingly capable of operating beyond isolated code generation. Repository-level benchmarks and agentic software engineering systems show that such agents can inspect existing code, navigate repositories, modify files, execute tools, and iteratively construct software patches \cite{chen2021codex,jimenez2024swebench,yang2024sweagent}. These capabilities make coding agents increasingly relevant to software maintenance, where the task is not to generate a program from scratch but to change an existing system while preserving behavior that should remain stable.

Software maintenance, however, involves more than locating code related to a request. A maintenance task may affect several connected artifacts, including controllers, service implementations, data-transfer objects, persistence mappings, configuration files, and regression tests. Maintenance and software-evolution research has long recognized that changes must be understood in relation to affected artifacts, dependencies, and accompanying test evidence \cite{iso14764,swanson1976maintenance,kretsou2021changeimpact,tian2021traceability,levin2017coevolution}. Consequently, an artifact may be relevant to a task without being an appropriate edit target. A controller may need to be inspected to understand an API route while the route itself must remain unchanged. Conversely, a regression test may not be mentioned in the user request but may still need to be modified to complete the maintenance task.

This distinction creates a problem for current coding-agent workflows. Repository search, retrieval, and tool-supported exploration can provide an agent with relevant information, but relevance alone does not determine how that information should be used. Retrieval can help answer which artifacts may contain useful evidence \cite{lewis2020rag}, while repository-level agents can independently explore and modify those artifacts \cite{yang2024sweagent}. Yet the agent must still infer a different set of task-specific decisions: which artifacts may be edited, which should only be inspected, which are required for task completion, and which should remain unchanged. In enterprise maintenance, these decisions are often guided by architectural conventions, maintenance practices, interface constraints, and testing expectations that human developers interpret as part of the task.

We argue that this gap concerns an operational boundary rather than repository access alone. Providing an agent with more context does not necessarily specify its permissible or required actions over that context. We therefore explore \emph{Boundary Contracts}, a machine-readable representation that assigns task-specific operational roles to software artifacts. A Boundary Contract distinguishes artifacts that are editable, inspect-only, required, or protected. It does not prescribe the implementation or provide the expected patch. Instead, it defines the operational space within which the coding agent performs repository exploration and maintenance.

The broader question is whether such explicit operational boundaries are useful enough to justify a distinct research direction in agentic software maintenance. This paper does not attempt to establish the general effectiveness of Boundary Contracts across repositories, programming languages, or coding-agent architectures. Instead, we conduct a pilot study in a real enterprise software repository to examine whether explicit artifact-role information produces a measurable change in coding-agent behavior and whether any observed effect appears to extend beyond file localization alone.

The pilot compares a coding agent operating with the original maintenance request against the same agent operating with the request plus a task-specific Boundary Contract. Across 40 repository-level maintenance tasks, the study examines both edit-scope decisions and resulting patch outcomes. In addition to the overall comparison, we analyze tasks in which both conditions reach at least one identical target file. This shared-file subset provides an exploratory probe of a central alternative explanation: that any observed benefit results only from helping the agent locate relevant files.

The purpose of this pilot is therefore not to present Boundary Contracts as a finished solution. Rather, it asks whether explicit operational artifact roles expose a sufficiently important and observable phenomenon to motivate deeper investigation. If coding-agent behavior can be improved by distinguishing what may be edited from what may only be inspected or must be preserved, a larger set of questions follows: how such boundaries should be generated, how their adequacy can be assessed, whether agents should be allowed to revise them during execution, and whether boundaries should remain advisory or be enforced by the execution environment. These questions motivate the broader research agenda developed later in this paper.

\section{The Missing Abstraction: From Relevance to Operational Role}
\label{sec:missing-abstraction}

Repository-level software maintenance requires an agent to reason about multiple artifacts rather than an isolated code fragment. Existing work provides several mechanisms for supporting this process. Repository-level benchmarks such as RepoBench and SWE-bench require models to operate over project context and resolve tasks in existing repositories \cite{liu2024repobench,jimenez2024swebench}. Agentic systems such as SWE-agent further allow models to search repositories, inspect files, modify code, and execute development tools \cite{yang2024sweagent}.

Recent approaches have also moved beyond repository access toward more structured support for multi-file changes. CodePlan, for example, formulates repository-level coding as a planning problem and uses dependency information, change-impact analysis, and iterative planning to coordinate interdependent edits \cite{bairi2024codeplan}. Such approaches demonstrate that repository-level coding requires more than exposing additional source files to an LLM: agents need mechanisms for reasoning about how changes propagate across software artifacts.

Retrieval and repository-level context provide another important source of support. Retrieval-augmented approaches can identify information that is potentially relevant to the current task \cite{lewis2020rag}, while repository-specific instruction files such as \texttt{AGENTS.md} attempt to provide coding agents with additional project knowledge and development guidance. Recent empirical evidence, however, shows that such context files do not generally improve task success and can increase inference cost, even though agents tend to follow the provided instructions \cite{gloaguen2026evaluating}. This suggests that providing more repository context or guidance is not by itself sufficient to determine effective maintenance behavior.

Software-maintenance research addresses complementary parts of the same problem. Bug localization seeks to identify program elements associated with a defect \cite{zhou2012buglocator,saha2013bluir}. Change impact analysis considers which software artifacts may be affected by a proposed change \cite{kretsou2021changeimpact}. Traceability research captures relationships among requirements, interfaces, implementation artifacts, and tests \cite{tian2021traceability}, while studies of test-code co-evolution show that changes to production code frequently require coordinated maintenance of test artifacts \cite{levin2017coevolution}. Together, these lines of work provide mechanisms for locating, relating, and coordinating artifacts involved in software maintenance.

However, identifying an artifact as relevant, related, or potentially affected does not determine what a coding agent should be allowed or expected to do with that artifact. An artifact may be relevant because it provides evidence needed to understand a task, while modifying it may violate an architectural or interface constraint. Conversely, an artifact that receives little emphasis in the original request---such as a regression test---may nevertheless be required for completing the maintenance task. Relevance therefore does not imply editability, and editability does not imply obligation.

We characterize this distinction as the difference between an artifact's \emph{informational relevance} and its \emph{operational role}. Informational relevance concerns whether an artifact is useful for understanding, locating, or planning a maintenance task. An operational role concerns what the agent is permitted or expected to do with that artifact during execution. For a particular maintenance task, an agent may therefore need to distinguish among at least four questions:

\begin{itemize}[leftmargin=*]
    \item Which artifacts may be modified?
    \item Which artifacts should be inspected but preserved?
    \item Which artifacts are required for task completion?
    \item Which artifacts must remain unchanged?
\end{itemize}

These questions overlap with repository planning, localization, impact analysis, and traceability, but they are not equivalent to them. Localization identifies where a relevant concern is likely to reside; impact analysis identifies artifacts that may be affected; traceability exposes relationships among artifacts; and planning approaches such as CodePlan organize interdependent changes. None of these signals alone specifies the task-specific \emph{permission} and \emph{obligation} associated with each artifact during agent execution. In particular, an artifact may participate in planning or provide useful evidence while still being an invalid edit target.

This distinction becomes more important as coding agents gain greater autonomy over repositories. Once an agent can directly modify software artifacts, uncertainty about artifact roles becomes an execution-control problem rather than only an information-retrieval or planning problem. Additional context may help determine what matters, but the agent must still decide whether each relevant artifact should be edited, inspected, required, or protected.

We therefore argue that agentic software maintenance benefits from an explicit representation of operational artifact roles. Such a representation should complement, rather than replace, retrieval, repository-level planning, localization, impact analysis, and traceability. These mechanisms can help determine which artifacts matter and how they are related; an operational boundary additionally specifies how those artifacts may participate in the maintenance task. The next section introduces Boundary Contracts as one possible realization of this missing abstraction.

\section{Boundary Contracts}
\label{sec:boundary-contracts}

We introduce \emph{Boundary Contracts} as a task-specific representation of the operational boundary within which a coding agent performs software maintenance. The purpose of a Boundary Contract is not to describe how a requested change should be implemented. Instead, it makes explicit how repository artifacts may participate in the maintenance task.

\subsection{Operational Artifact Roles}

A Boundary Contract assigns operational roles to artifacts associated with a maintenance task. In the current design, these roles capture four basic distinctions:

\begin{itemize}[leftmargin=*]
    \item \textbf{Editable}: artifacts that the agent is permitted to modify;
    \item \textbf{Inspect-only}: artifacts that may provide evidence but should remain unchanged;
    \item \textbf{Required}: artifacts, or artifact types, expected to participate in completing the task; and
    \item \textbf{Protected}: artifacts that should not be modified for the requested change.
\end{itemize}

The contract can express these roles through fields such as
\texttt{allowed\_read\_paths},
\texttt{allowed\_edit\_paths},
\texttt{must\_edit},
\texttt{inspect\_only},
\texttt{must\_not\_edit}, and task-specific
\texttt{violation\_rules}. These fields distinguish relevance from permission and obligation. For example, an API controller may be relevant for understanding a request while remaining inspect-only, whereas a regression test may be required even if it is not explicitly named in the request.

The central idea is therefore that a Boundary Contract constrains the \emph{operational space} of maintenance rather than prescribing the implementation. It does not contain the expected code diff, implementation statements, or final patch. The coding agent remains responsible for repository exploration, implementation reasoning, and patch generation.

\subsection{Constructing a Task-Specific Boundary}

Boundary Contracts are generated from reusable maintenance knowledge rather than manually authored as task-specific gold answers. Before individual maintenance requests are processed, repository information is organized into a maintenance knowledge base containing repository inventory, maintenance units, artifact relationships, recurring change-type schemas, and maintenance rules.

A \emph{maintenance unit} represents a set of artifacts that commonly participate together in maintenance. Depending on the system, such a unit may include API specifications, controllers, request and response DTOs, service implementations, persistence mappings, configuration files, and tests. A \emph{change-type schema} captures a recurring maintenance pattern, such as adding a response field, changing a validation rule, modifying an SQL condition, or changing calculation logic. Each schema provides reusable expectations about artifact roles for that type of change.

When a maintenance request arrives, a proposal agent retrieves candidate maintenance units and identifies candidate change types. It then combines these candidates with the corresponding maintenance rules to construct a task-specific boundary pack. In our implementation, this pack is materialized before repository modification as \texttt{CANDIDATES.yaml}.

Importantly, the boundary pack is generated independently of the gold answer used for evaluation. The reusable maintenance knowledge is prepared separately from the task-specific gold evaluation records and is not derived from generated patches. The boundary pack does not contain the gold edit-target set, expected code changes, or post-hoc evaluation labels. Its purpose is to provide operational guidance before the coding agent begins modifying the repository.

\subsection{Boundary-Informed Agent Execution}

\begin{figure}[t]
    \centering
    \includegraphics[width=\columnwidth]{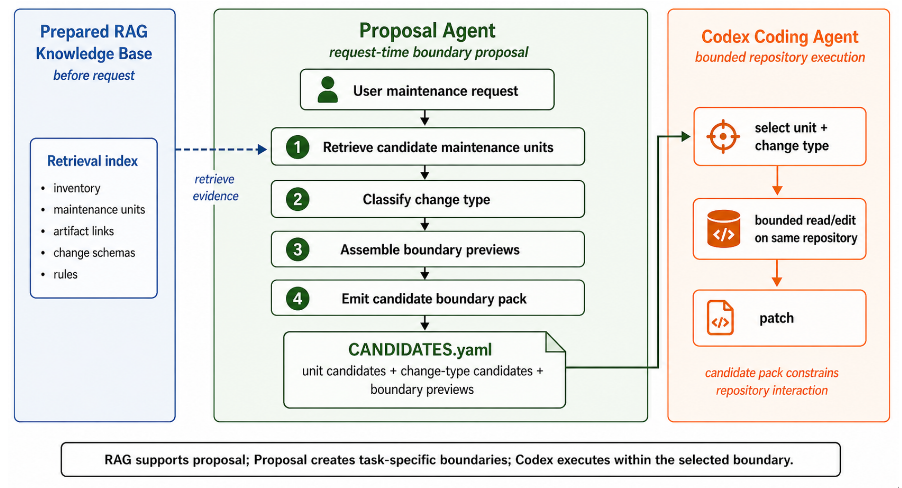}
    \caption{Boundary-informed maintenance workflow. A proposal agent
    constructs task-specific operational boundaries before the coding agent
    performs repository maintenance.}
    \label{fig:boundary-workflow}
\end{figure}

Figure~\ref{fig:boundary-workflow} illustrates the resulting workflow. 

%insert picture

The proposal agent and coding agent have distinct responsibilities. The proposal agent constructs candidate operational boundaries from reusable maintenance knowledge, while the coding agent receives both the original maintenance request and the generated boundary pack.

The coding agent still selects the candidate maintenance unit and change type that best fit the request, inspects repository artifacts, determines the implementation, and generates the patch. The Boundary Contract therefore does not replace the agent's reasoning. It reduces one particular form of uncertainty by making artifact roles explicit before modification begins.

This separation also distinguishes Boundary Contracts from ordinary repository retrieval. Retrieval determines which artifacts should be made available as potentially useful context. A Boundary Contract additionally states how those artifacts are expected to participate in execution. In this sense, the contract acts as an operational interface between reusable maintenance knowledge and the autonomous actions of the coding agent.

The design deliberately leaves an important question unresolved. If the supplied boundary is incomplete or incorrect, the coding agent may discover during execution that an inspect-only or excluded artifact appears necessary for completing the task. Whether the agent should obey, challenge, or revise such a boundary is not determined by the current mechanism. This limitation becomes a central research question in Section~\ref{sec:research-agenda}.

\section{Pilot Study: Is the Idea Worth Pursuing?}
\label{sec:pilot-study}

The purpose of this pilot is to examine whether explicit operational artifact roles have a measurable effect on coding-agent behavior before undertaking a broader evaluation across repositories, agents, and maintenance settings. We therefore use a controlled comparison in which the maintenance tasks, repository, coding agent, and execution procedure are held constant, while only the availability of Boundary Contract information is varied. This follows the general principle of controlled empirical software engineering studies in which the treatment of interest is varied while other experimental conditions are kept stable \cite{Kitchenham2002Preliminary,Wohlin2012Experimentation}.

\subsection{Pilot Questions}

The pilot is guided by two questions:

\begin{description}[leftmargin=0pt,labelindent=0pt]
    \item[\textbf{PQ1.}] Does explicit artifact-role information change the maintenance scope selected by a coding agent?
    \item[\textbf{PQ2.}] Is there initial evidence that the effect of explicit artifact roles extends beyond file localization to the resulting patch outcome?
\end{description}

PQ1 examines whether the agent is more likely to modify expected artifacts while avoiding unnecessary or protected edits. PQ2 addresses a stronger alternative explanation. If Boundary Contracts only help the agent locate relevant files, any improvement may primarily reflect better localization rather than a distinct operational-boundary effect. We therefore examine both overall patch outcomes and a subset of tasks in which the baseline and boundary-informed conditions reach at least one identical file.

\subsection{Study Context and Maintenance Tasks}

The study uses the source repository of a real enterprise information system containing more than 2,900 Java and XML artifacts. The repository was selected because maintenance requests in this environment commonly span multiple software artifacts and therefore require decisions about what should be modified, inspected, coordinated, or preserved.

We construct eight recurring maintenance-task groups: response-field addition, request-driven condition addition, SQL-condition change, domain-code mapping, validation-rule addition, response-naming change, calculation-logic change, and no-routing-change preservation. These groups are used as sampling strata rather than as a proposed universal taxonomy of software maintenance. Each group contains five task instances, producing 40 repository-level maintenance tasks in total. The tasks span 34 distinct expected maintenance units, reducing repeated evaluation of the same program location under different wording.

Each task contains a natural-language maintenance request. A separate gold record specifies the expected maintenance outcome for evaluation, including expected edit targets, protected targets, required behaviors, and required test evidence. Gold records are used only after patch generation and are not provided to the coding agent.

\subsection{Experimental Conditions}

Each of the 40 tasks is executed under two conditions.

\begin{itemize}[leftmargin=*]
    \item \textbf{Baseline}: the coding agent receives the natural-language maintenance request and has access to the repository and its normal exploration tools.
    \item \textbf{Boundary-informed}: the same coding agent receives the same request and repository access, together with the task-specific candidate Boundary Contract described in Section~\ref{sec:boundary-contracts}.
\end{itemize}

The baseline is therefore not prevented from searching the repository or discovering relevant files. It must independently infer the maintenance unit, edit scope, companion artifacts, and artifacts that should remain unchanged. The boundary-informed condition differs only in receiving explicit candidate artifact roles before modification begins.

All executions use the same Codex-based coding-agent environment, repository access, tool permissions, task-prompt structure, and execution procedure. Each task is executed once under each condition, producing 80 generated patches. The pilot is therefore a within-task controlled comparison rather than a comparison between different models or agent frameworks.

\subsection{Measures}

We examine two complementary aspects of the generated patches.

\textbf{Maintenance-scope selection} measures whether the agent modifies the expected repository artifacts while avoiding unnecessary edits. We record expected targets reached, omitted expected edits, extra edits, and modifications to protected artifacts. These counts are summarized using task-level and aggregate precision, recall, and F1.

\textbf{Patch outcome} measures whether the selected modifications satisfy the intended maintenance behavior. For each task, predefined evaluation records identify required behavior items and required test evidence. Patch review records the number of satisfied behavior items, blocking defects, regression-risk findings, and satisfied test requirements. A patch is considered strictly acceptable only when all required behavior is satisfied, no blocking defect or regression-risk finding remains, and all required test evidence is present.

Patch assessment uses a fixed count-based coding schema. An LLM-assisted review produces an initial coding draft, which is manually checked and corrected before aggregation. The evaluator does not assign an unrestricted holistic quality score; the final measures are derived from predefined task-specific evaluation fields.

These measures are not intended to establish general effectiveness from a single pilot. Their purpose is to determine whether the proposed operational-boundary abstraction produces a sufficiently clear behavioral signal to warrant broader investigation.

\section{Pilot Evidence: More Than File Localization?}
\label{sec:pilot-evidence}

The pilot provides an initial positive signal for both maintenance-scope selection and patch outcome. However, the purpose of this analysis is not simply to report that the boundary-informed condition performs better. A more important question is whether the observed improvement can be explained only by helping the agent reach relevant files.

\subsection{Overall Pilot Results}

Table~\ref{tab:pilot-results} summarizes the main results. For maintenance-scope selection, macro F1 increases from 0.440 in the baseline condition to 0.682 in the boundary-informed condition, while micro F1 increases from 0.438 to 0.726. The boundary-informed condition reaches more expected edit targets and makes fewer unnecessary edits: omitted expected edits decrease from 62 to 34, and extra edits decrease from 56 to 22.

Patch outcomes also improve. Required-behavior coverage increases from 0.429 to 0.673, blocking defects decrease from 52 to 26, and required test evidence satisfied increases from 11 of 40 to 26 of 40 tasks. Under the strict acceptance criterion, which requires all required behavior and test evidence to be satisfied with no blocking defect or regression-risk finding, acceptable patches increase from 9 of 40 to 18 of 40.

\begin{table}[t]
\centering
\caption{Selected results from the 40-task pilot.}
\label{tab:pilot-results}
\small
\begin{tabular}{lrr}
\toprule
\textbf{Metric} & \textbf{Baseline} & \textbf{Boundary} \\
\midrule
Macro selection F1           & 0.440 & 0.682 \\
Micro selection F1           & 0.438 & 0.726 \\
Omitted expected edits       & 62    & 34    \\
Extra edits                  & 56    & 22    \\
Behavior coverage            & 0.429 & 0.673 \\
Blocking defects             & 52    & 26    \\
Required tests satisfied     & 11/40 & 26/40 \\
Strict acceptable patches    & 9/40  & 18/40 \\
\bottomrule
\end{tabular}
\end{table}

These results answer PQ1 positively within the pilot setting: explicit artifact-role information is associated with a substantial change in the maintenance scope selected by the coding agent. The agent covers more expected artifacts while making fewer edits outside the expected scope.

\subsection{Could This Be Only a Localization Effect?}

The overall comparison alone does not establish that Boundary Contracts contribute anything beyond localization. The boundary-informed condition receives candidate maintenance units and artifact roles, and this additional information can make it easier to reach relevant parts of the repository. If improved localization entirely explains the result, the proposed abstraction would be closer to structured retrieval assistance than to a distinct form of operational control.

To probe this alternative explanation, we examine the 21 tasks in which both conditions modify at least one identical file. This subset does not remove all localization differences, but it restricts the analysis to tasks where the two executions share at least part of their selected repository scope.

Table~\ref{tab:same-target-results} shows that differences remain within this subset. Macro selection F1 increases from 0.652 to 0.786, while extra edits decrease from 29 to 9. Required-behavior coverage increases from 0.673 to 0.750, and acceptable patches increase from 7 of 21 to 13 of 21. Blocking defects decrease from 19 to 11, and strict acceptable patches increase from 7 of 21 to 11 of 21.

\begin{table}[t]
\centering
\caption{Selected results for the 21 tasks where both conditions modified at least one identical file.}
\label{tab:same-target-results}
\small
\begin{tabular}{lrr}
\toprule
\textbf{Metric} & \textbf{Baseline} & \textbf{Boundary} \\
\midrule
Macro selection F1        & 0.652 & 0.786 \\
Extra edits               & 29    & 9     \\
Behavior coverage         & 0.673 & 0.750 \\
Blocking defects          & 19    & 11    \\
Acceptable patches        & 7/21  & 13/21 \\
Strict acceptable patches & 7/21  & 11/21 \\
\bottomrule
\end{tabular}
\end{table}

The shared-file subset is informative because the boundary-informed condition continues to show differences even when both executions modify at least one common repository artifact. In particular, the reduction in extra edits suggests that explicit artifact roles may influence not only where the agent searches, but also how broadly it modifies the repository after reaching relevant artifacts.

At the same time, this analysis should not be interpreted as a clean causal separation between localization and operational-boundary effects. The two conditions may still inspect different additional files, select different companion artifacts, or follow different reasoning trajectories. A stronger study would need to control localization directly---for example, by supplying identical retrieved context to both conditions and varying only the operational roles assigned to those artifacts.

\subsection{Failures Are Also Informative}

Boundary information does not guarantee a successful patch. In the pilot, some boundary-informed executions still fail to implement required behavior or provide required test evidence. In particular, failures occur when the supplied boundary is insufficient or too narrow, or when the agent reaches the relevant scope but still fails to synthesize the required implementation logic.

These failures expose an important limitation of the current design. Boundary Contracts can reduce uncertainty about maintenance scope, but they introduce a new dependency on the adequacy of the boundary itself. A correct boundary may improve the conditions under which implementation occurs; an incomplete or incorrect boundary may instead constrain the agent in the wrong direction.

The pilot therefore supports a deliberately limited conclusion. It provides initial evidence that explicit operational artifact roles can affect both maintenance-scope selection and patch outcomes, and that the observed effect may extend beyond file localization alone. It does not establish general effectiveness or isolate the mechanism completely. Instead, the remaining ambiguity motivates the deeper research questions considered in the next section.

\section{The Harder Problem: Who Controls the Boundary?}
\label{sec:research-agenda}

The pilot suggests that explicit artifact roles can influence how a coding agent performs repository-level maintenance. However, this result creates a harder problem. If an agent's actions are constrained by an operational boundary, the quality of maintenance increasingly depends on whether that boundary is itself appropriate.

This issue is already visible in the pilot. Boundary-informed executions still fail when the provided scope is too narrow, when a required companion artifact is missing, or when the agent reaches the relevant artifacts but cannot synthesize the required implementation. Boundary Contracts therefore move part of the uncertainty in agentic maintenance from \emph{where and how the agent should act} to \emph{how the operational boundary itself should be constructed and controlled}.

We see five questions as particularly important for developing explicit operational boundaries beyond the current pilot.

\subsection{How Should Boundaries Be Generated?}

The current prototype constructs candidate boundaries from prepared maintenance knowledge, including maintenance units, artifact relationships, recurring change types, and maintenance rules. This demonstrates one feasible mechanism, but does not establish that this is the best or most general way to produce boundaries.

A larger study should distinguish the quality of boundary generation from the ability of an agent to use a correct boundary. These are separate problems. An agent may benefit greatly from accurate artifact roles while a practical system may still fail if those roles cannot be generated reliably.

Future evaluations should therefore measure boundary quality directly. Candidate boundaries can be compared with independently defined maintenance expectations to measure whether editable, required, inspect-only, and protected roles are assigned correctly. This would make it possible to study how errors in boundary generation propagate into patch outcomes.

\subsection{How Much Boundary Error Can an Agent Tolerate?}

Operational boundaries will rarely be perfect. They may omit a required artifact, classify an editable artifact as inspect-only, include unnecessary editable artifacts, or fail to protect an interface that should remain stable.

This raises a robustness question that is not addressed by the current pilot:

\begin{quote}
How does agent behavior change as the operational boundary becomes incomplete or incorrect?
\end{quote}

A controlled study could deliberately perturb otherwise valid Boundary Contracts. For example, required artifacts could be removed, protected artifacts could be exposed as editable, or irrelevant artifacts could be added to the permitted scope. Measuring the resulting changes in task success, unnecessary edits, boundary violations, and recovery behavior would reveal whether Boundary Contracts fail gradually or whether particular boundary errors create sharp failure modes.

Such an experiment would also help determine which artifact roles matter most. The current pilot evaluates the contract as a combined treatment. It does not isolate whether improvements arise primarily from identifying required artifacts, protecting stable artifacts, restricting editable scope, or distinguishing inspect-only evidence. An ablation study across these roles is therefore necessary before strong claims can be made about the mechanism.

\subsection{Should a Coding Agent Be Allowed to Revise Its Boundary?}

A static boundary creates a tension between control and autonomy. An agent may discover during repository exploration that an artifact outside the original editable scope is necessary for completing the requested change. Strictly enforcing the original contract may then prevent a correct implementation, while allowing the agent to ignore the contract may undermine the purpose of having an operational boundary.

This leads to a central design question:

\begin{quote}
Should a coding agent be allowed to override a maintenance boundary that it believes is incomplete?
\end{quote}

Several designs are possible. An agent could be prohibited from crossing the boundary; it could request an explicit boundary extension from another agent or human reviewer; or it could revise the boundary autonomously while recording the justification for the change. These alternatives represent different trade-offs between autonomy, safety, and development efficiency.

This suggests that operational boundaries may need to become dynamic rather than static. Instead of treating a Boundary Contract as a fixed input, future systems could treat it as a maintained execution artifact whose revisions are proposed, justified, approved, and recorded during the maintenance process.

\subsection{Should Boundaries Be Advisory or Enforced?}

In the current pilot, Boundary Contracts provide operational guidance to the coding agent. The agent is instructed how artifacts should be treated, but the contract is primarily communicated through the agent's task context.

A stronger form of boundary control would move enforcement into the execution environment. Repository tools could, for example, permit reads from inspect-only artifacts while rejecting edit operations, prevent modifications to protected paths, or require explicit authorization before expanding the editable scope.

These two designs answer different questions. Prompt-level boundaries test whether an agent can reason effectively when artifact roles are made explicit. Runtime-enforced boundaries test whether those roles can serve as actual controls over autonomous software modification.

The distinction matters for enterprise use. If a boundary is intended only to improve maintenance quality, advisory guidance may be sufficient. If it is also intended to prevent undesirable changes, the boundary may need to become an enforceable policy rather than merely additional context.

\subsection{Does the Need for Boundaries Generalize?}

Finally, the current evidence comes from one enterprise repository and one Codex-based coding-agent configuration. It does not establish whether the same phenomenon appears across different repository structures, programming languages, agent architectures, or model capabilities.

A broader evaluation should therefore separate two questions. First, does providing explicit operational roles consistently improve maintenance behavior across systems? Second, under what conditions do boundaries provide the greatest benefit?

For example, stronger agents may infer maintenance roles sufficiently well in highly regular repositories, reducing the value of an explicit contract. Conversely, large or heterogeneous enterprise systems with substantial architectural conventions may benefit more from explicit operational boundaries. The value of Boundary Contracts may therefore depend on both agent capability and repository complexity.

A full evaluation should combine multiple repositories, maintenance-task families, and coding agents with controlled localization conditions. In particular, supplying identical retrieved artifacts to both baseline and boundary-informed agents while varying only the operational roles would provide a stronger test of whether the effect observed in this pilot is genuinely attributable to boundary information.

Taken together, these questions shift the research problem beyond whether Boundary Contracts improve a particular coding agent. The broader challenge is how operational authority over a software repository should be represented, generated, validated, revised, and enforced when maintenance is performed by autonomous or semi-autonomous agents. The pilot provides an initial reason to investigate that question, but answering it requires substantially broader evidence than the present study provides.

\section{Limitations and Conclusion}
\label{sec:conclusion}

This study is deliberately limited in scope. The pilot uses one enterprise repository, 40 maintenance tasks across eight recurring task groups, and a single Codex-based coding-agent configuration. The observed effects may therefore depend on the repository structure, maintenance conventions, task distribution, or capabilities of the evaluated agent. In addition, patch outcomes are assessed using predefined patch-based evaluation records rather than full execution of the repository's test suite. The study consequently does not establish runtime correctness of the generated patches. Each task was also executed only once per condition, so the pilot does not estimate run-to-run variability in coding-agent behavior.

A further limitation is that the effectiveness of a Boundary Contract depends on the quality of the boundary supplied to the agent. The current pilot evaluates the combined effect of candidate artifact roles and does not independently measure boundary-generation accuracy or isolate the contribution of editable, required, inspect-only, and protected roles. Nor does the shared-file analysis completely remove localization as a possible explanation for the observed improvements. In addition, patch coding was manually verified by an evaluator, but the pilot does not include independent double coding or an inter-rater agreement analysis. These limitations restrict the claims that can be made from the present evidence.

Within these constraints, however, the pilot provides an initial signal that explicit operational artifact roles are worth investigating. Across the evaluated tasks, the boundary-informed agent modifies maintenance scope more selectively and produces stronger patch outcomes than the baseline. Importantly, differences remain in the subset where both conditions reach at least one identical file, suggesting that the effect may extend beyond file localization alone.

The contribution of this pilot is therefore not a claim that Boundary Contracts solve agentic software maintenance. Rather, it identifies a broader problem: repository access and relevant context do not by themselves specify the operational authority of a coding agent over the artifacts it encounters. Boundary Contracts provide one concrete abstraction for making that authority explicit.

The next step is to determine how such boundaries should be generated, validated, revised, and enforced across different repositories and agent architectures. Ultimately, the question is not only whether coding agents can autonomously navigate and modify software repositories, but how their autonomy over those repositories should be bounded.

\bibliographystyle{ACM-Reference-Format}
\bibliography{references}

\end{document}